\documentclass[namedreferences]{SolarPhysics}
\usepackage[optionalrh]{spr-sola-addons} 
\usepackage{graphicx}                    
\usepackage{color}                       
\usepackage{url}                         

\begin{document}

\begin{article}

\begin{opening}
\title{Oscillation Power in Sunspots and Quiet Sun from Hankel Analysis Performed on SDO/HMI and SDO/AIA Data}

\author{S\'ebastien Couvidat$^{1}$}

\institute{$^{1}$ W.W. Hansen Experimental Physics Laboratory, Stanford University, CA 94305, USA\\}

\date{Received: 31 May 2012, Revised: 30 July 2012}

\runningauthor{S. Couvidat}
\runningtitle{Oscillation Power in Sunspots and Quiet Sun from Hankel Analysis}

\begin{abstract}
The {\it Helioseismic and Magnetic Imager} (HMI) and the {\it Atmospheric Imaging Assembly} (AIA) instruments onboard the \textit{Solar Dynamics Observatory} satellite produce Doppler velocity and continuum intensity at 6173 \AA\ as well as intensity maps at 1600 \AA\ and 1700 \AA, which can be used for helioseismic studies at different heights in the solar photosphere.  
We perform a Hankel-Fourier analysis in an annulus centered around sunspots or quiet-Sun regions, to estimate the change in power of waves crossing these regions of interest.
We find that there is a dependence of power-reduction coefficients $\alpha$ on measurement height in the photosphere: Sunspots reduce the power of outgoing waves with frequencies $\nu$ lower than $\nu \approx 4.5$ mHz at all heights, but enhance the power of acoustic waves in the range $\nu \approx 4.5-5.5$ mHz toward chromospheric heights, which is likely the signature of acoustic glories (halos). Maximum power reduction seems to occur near the continuum level and to decrease with altitude. Sunspots also impact the frequencies of outgoing waves in an altitude-dependent fashion.
The quiet Sun is shown to behave like a strong power reducer for outgoing $f$ and $p$-modes at the continuum level, with a power reduction $\alpha \approx 15-20\%$, and like a weak power enhancer for $p$-modes higher in the atmosphere. It is speculated that the surprising power reduction at the continuum level is related to granulation. In Doppler-velocity data, and unlike in intensity data, the quiet Sun behaves like a strong power reducer for granular flows.
\end{abstract}


\end{opening}

\section{Introduction}

The {\it Helioseismic and Magnetic Imager} instrument (HMI; Schou {\it et al.}, 2012) onboard the {\it Solar Dynamics Observatory} satellite (SDO) measures, in the Fe {\sc i} line at $6173$ \AA\ ({\it e.g.} Dravins, Lindegren, and Nordlund, 1981; Norton {\it et al.}, 2006), the motion of the solar photosphere. HMI samples the iron line at six wavelengths symmetrical around the line center at rest.
Line-of-sight (LOS) observables are produced every 45 s, which are Doppler velocity, LOS magnetic-field strength, Fe {\sc i} line width, line depth, and continuum intensity. These are calculated from an algorithm based on the one used for the MDI instrument ({\it Michelson Doppler Imager}; Scherrer {\it et al.}, 1995). A brief description of this algorithm is given in Couvidat {\it et al.} (2012). The {\it Atmospheric Imaging Assembly} instrument (AIA; Lemen {\it et al.}, 2012) images the Sun in ten EUV and UV wavelength channels.
It has conclusively been shown that two channels, at 1600 \AA\ and 1700 \AA, can be used for helioseismic studies ({\it e.g.} Hill {\it et al.}, 2011; Howe {\it et al.}, 2011, 2012; Rajaguru {\it et al.}, 2012).

The formation height of the signal in these two bands is $\approx 430$ and $\approx 360$ km, respectively, based on the study by Fossum and Carlsson (2005) for the TRACE mission. Images in these bands are taken at a 24-s cadence.
The formation height of the HMI Doppler velocity is $\approx 150$ km (Fleck, Couvidat, and Straus, 2011), while the formation height of the core of the Fe {\sc i} line is $\approx 302$ km (Norton {\it et al.}, 2006).
Therefore, by combining HMI and AIA measurements, the entire photospheric layer and the lower chromosphere can be sampled (AIA UV bands are sensitive to heights up to $\approx 800$ km). This brings forth new opportunities to study this layer with the additional height information.

In this paper, a Hankel-Fourier transform (Braun, Duvall, and LaBonte, 1987) is applied to sunspot data, to measure the power reduction and/or enhancement in the $f$ and $p$-oscillation modes crossing an active region. Braun, Duvall, and LaBonte (1987) established that sunspots ``absorb'' acoustic power, and this power reduction has been observed at different wavelengths, but no attempt has been made at measuring this reduction simultaneously at different atmospheric heights. Instead, different groups used individual datasets. For instance, Braun, Duvall, and LaBonte (1987, 1988) used Dopplergrams of the Fe {\sc i} line at 8688 \AA\ formed in higher photospheric layers, Bogdan {\it et al.} (1993) used Dopplergrams of the Fe {\sc i} line at 5576 \AA\ formed in the lower photosphere, Braun (1995) used Ca {\sc ii} K line intensity images formed in the chromosphere, and Zhang (1997) used intensity images at 3940 \AA\ sensitive to the middle and upper photosphere. Moreover, MDI and HMI continuum-intensity data appear not to have been used in any publication related to the Hankel-Fourier analysis. Even though Braun and Fan (1998) did use MDI intensities to measure meridional circulation, no power-reduction analysis was performed.

The difference, if any, in power reduction at different heights should provide additional information regarding the physical mechanisms at play. Indeed, the complex interplay between magnetic field and plasma in the photosphere deeply impacts wave-propagation properties, probably in a height-dependent fashion. Whether the waves are partly running or completely standing, whether the mechanism responsible for power reduction operates at a given height or across a significant altitude or depth range, whether specific frequencies or wavelengths are affected at different heights, are but a few questions that can be investigated by Hankel-Fourier transforms using contemporaneous HMI and AIA data.

Amongst all of the magnetic phenomena impacting acoustic-gravity waves, one of the earliest known is the reduction in amplitude of $p$-modes by active regions (Woods and Cram, 1981). Active regions also convert the acoustic and surface gravity waves into magneto-acoustic-gravity (MAG) waves ({\it e.g.} Spruit and Bogdan, 1992; Cally and Bogdan, 1993; Crouch {\it et al.}, 2005; Khomenko and Collados, 2006) that propagate higher in the atmosphere and deeper in subphotospheric layers. Magnetic fields inhibit granulation, resulting in a lower acoustic-wave emissivity in sunspots ({\it e.g.} Parchevsky and Kosovichev, 2007; Hanasoge {\it et al.}, 2008), and lower the acoustic cut-off frequency ({\it e.g.} De Pontieu, Erdelyi, and James, 2004), allowing otherwise standing waves to travel upwards. Finally, magnetic fields modify acoustic-wave frequencies.

Even though the photosphere is stably stratified, overshoot from the convective zone produces a clear granulation and supergranulation signal in photospheric measurements that varies with altitude. Indeed, as pointed out by Howe {\it et al.} (2011), AIA 1600 and 1700 \AA\ intensity maps are much less contaminated by granulation noise than the HMI continuum. Therefore, noise properties are also height dependent, providing yet another rationale to performing helioseismic studies simultaneously on the different SDO observables.

Several local-helioseismology techniques exist to study wave properties and infer the subphotospheric structure and dynamics ({\it e.g.} time-distance analysis, Duvall {\it et al.}, 1993; acoustic holography, Lindsey and Braun, 1998; ring-diagram analysis, Hill, 1988). These techniques complement or supersede the somewhat simpler Hankel-Fourier transform, recently fallen out of favor. Each one of these techniques provides a fragmentary picture of how strong magnetic fields in sunspots affect waves, and has been applied to data from various instruments, whose signals are not necessarily formed at the same height in the solar atmosphere. Here, the Hankel-Fourier transform is performed on signal from outside the sunspots ({\it i.e.} not contaminated by strong fields) to facilitate the interpretation of its results, and we use simultaneous measurements at four different altitudes. We study both sunspots and quiet Sun regions.
In Section 1 we briefly remind the reader of the Hankel-Fourier analysis. In Section 2 we describe the data used. In Section 3 we present our results, and we conclude in Section 4.

\section{Hankel-Fourier Transform}

The Hankel-Fourier decomposition is extensively described in Braun, Duvall, and LaBonte (1987, 1988), and Braun (1995).
Therefore, only a brief description will be given here. In polar coordinates $(r,\theta)$ and as a function of time $t$, the wave field $\phi$ can be decomposed into components of the form
\begin{equation}
\phi_m(r,\theta,t) = e^{i(m \theta + \omega t)} [ A_m(k,\omega) H^{(1)}_m(kr) + B_m(k,\omega) H^{(2)}_m(kr)]
\end{equation}
where $m$ is the azimuthal order, $k$ is the horizontal wavenumber, $\omega=2 \pi \nu$ is the cyclical frequency ($\nu$ is the temporal frequency), and $H^{(1)}_m$ and $H^{(2)}_m$ are Hankel functions of the first and second kind, respectively. Coefficients $A_m$ and $B_m$ are complex amplitudes of the radially ingoing and radially outgoing waves, respectively.
As mentioned in Bogdan {\it et al.} (1993), Hankel functions are an approximation to the Legendre functions, valid for $\ell \gg m$, where $\ell$ is the angular degree (Braun, 1995). Here we only compute the power of ingoing and outgoing waves, {\it i.e.} $|A_m|^2$ and $|B_m|^2$, and neglect the phases of these coefficients. Indeed, phase measurement is a more complicated matter and will be addressed at a later time.
A measure, in an annulus centered on $r=0$, of the reduction in power of outgoing waves $\alpha_m(k,\omega)$, is obtained by
\begin{equation}
\alpha_m(k,\omega) = 1-\frac{|B_m(k,\omega)|^2-N_m(k,\omega)}{|A_m(k,\omega)|^2-N_m(k,\omega)}
\end{equation}
where $N_m(k,\omega)$ is an estimate of the background power at $k$ and $\omega$ for azimuthal order $m$. We are primarily interested in the power of oscillation modes. Therefore we subtract from the power spectra a background power resulting, mainly, from granulation and supergranulation.

Quantity $\alpha_m(k,\omega)$ is traditionally referred to as absorption coefficient in the literature. However, this coefficient includes the effects of not only power absorption, but also emissivity reduction arising inside sunspots (Chou {\it et al.}, 2009b). The mechanisms responsible for these two effects are detailed in Chou {\it et al.} (2009a). Therefore, it is more appropriate to refer to $\alpha_m(k,\omega)$ as a power-reduction coefficient (\mbox{T.L.} Duvall, Jr., 2012, private communication). 

To increase the signal-to-noise (S/N) ratio, an $m$-averaged outgoing-wave power-reduction coefficient $\alpha(k,\omega)$ is calculated, based on the sum of individual $m$ power spectra (with or without background subtracted). For instance, with background subtracted, $\alpha(k,\omega)$ can be defined as
\begin{equation}
\alpha(k,\omega) = 1-\frac{(\sum_m |B_m(k,\omega)|^2)-N(k,\omega)}{(\sum_m |A_m(k,\omega)|^2)-N(k,\omega)}
\end{equation}
where $N(k,\omega)$ is an estimate of the background calculated on the $m$-averaged power spectra. The same background is subtracted from ingoing and outgoing spectra. The calculation of $\alpha(k,\omega)$ is somewhat sensitive to the way this background is removed, especially for the HMI continuum intensity where granulation power is the highest. Therefore raw $\alpha(k,\omega)$ values with no background subtraction are also presented here. Two techniques of background measurement are applied: one fitting a single background $N$ to the $m$-averaged power spectra (based on Braun, Duvall, and LaBonte, 1988), and one where a separate background power $N_m$ is fitted to each individual $m$ power spectrum (based on Braun, 1995). Both methods give similar results. In both cases, the logarithm of the power spectral density (PSD) is fitted by a 5th-order polynomial as a function of $\nu$, at each $k$. The $f$ and $p_1$ to $p_9$-mode ridges (oscillation modes with radial order $n=0$ to $n=9$) are excluded from the fit.

To further improve the S/N ratio, we may integrate the PSDs over $\omega$, or $k$, or both, prior to calculating power-reduction coefficients. Thus, we define three quantities (here with the first method of background-power removal), $\alpha(k)$, $\alpha(\omega)$, and $\alpha$, where
\begin{equation}
\alpha(k) = 1-\frac{\sum_{\omega}((\sum_m |B_m(k,\omega)|^2)-N(k,\omega))}{\sum_{\omega}((\sum_m |A_m(k,\omega)|^2)-N(k,\omega))},
\end{equation}
\begin{equation}
\alpha(\omega) = 1-\frac{\sum_{k}((\sum_m |B_m(k,\omega)|^2)-N(k,\omega))}{\sum_{k}((\sum_m |A_m(k,\omega)|^2)-N(k,\omega))},
\end{equation}
and
\begin{equation}
\alpha = 1-\frac{\sum_k \sum_{\omega}((\sum_m |B_m(k,\omega)|^2)-N(k,\omega))}{\sum_k \sum_{\omega}((\sum_m |A_m(k,\omega)|^2)-N(k,\omega))}.
\end{equation}
A mask is usually applied to the PSDs, to only select power within the oscillation-mode ridges ({\it e.g.} Braun, Duvall, and LaBonte, 1988). Here, the mask only keeps the modes with $n \le 9$ and above 2 mHz.

The polar coordinate system is centered at a region of interest, {\it e.g.} the center of a sunspot, and the ingoing and outgoing-wave decomposition is calculated in an annulus with inner radius $R_\mathrm{min}$ selected to exclude this sunspot.
For instance, Doppler-velocity measurements in strong magnetic fields are known to be unreliable ({\it e.g.} Rajaguru, Wachter, and Hasan, 2006), which makes it a better option to only use signal from outside strong-field areas.

\section{Data Used}

\begin{figure}
\centering
\includegraphics[width=\textwidth]{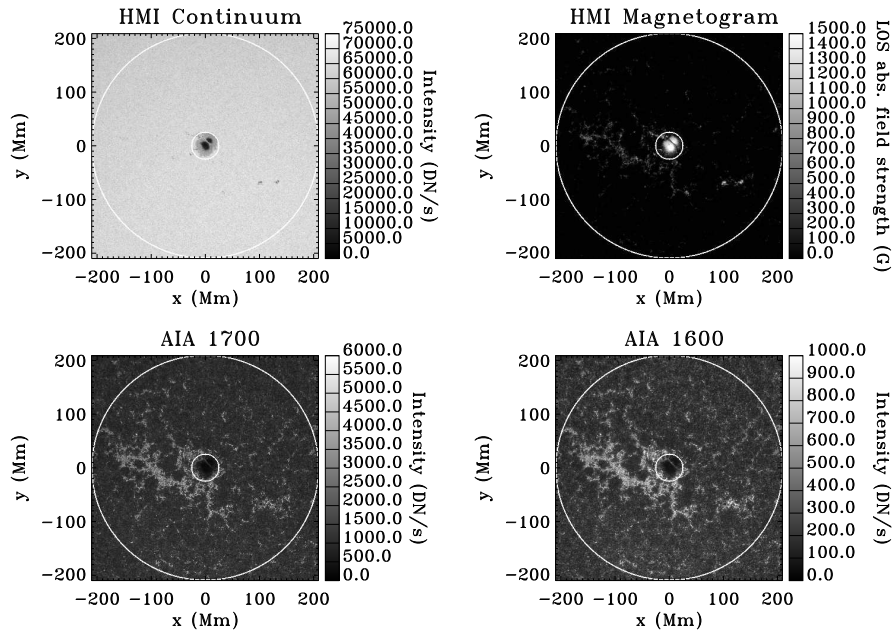}
\caption{Snapshot of a solar region centered on sunspot NOAA 11289 from the HMI continuum intensity (upper left panel), absolute value of the HMI LOS magnetic-field strength (upper right panel), AIA 1700 \AA\ intensity (lower left panel), and AIA 1600 \AA\ intensity (lower right panel). The annulus inside which the Hankel-Fourier transform is calculated is defined by the two white circles.}
\label{firstfig}       
\end{figure}

We use HMI continuum intensity, HMI LOS Doppler velocity, AIA 1700 \AA, and AIA 1600 \AA\ intensity maps. We also occasionally use HMI line-core intensity, formed by subtracting the Fe {\sc i} line depth from the continuum intensity. The LOS magnetic flux in sunspots and quiet-Sun regions is estimated from HMI LOS magnetic-field strength (magnetogram).
Each datacube is tracked for four days at the Carrington rotation rate using the {\sf mtrack} software. A Postel projection is performed at each time step.
The spatial resolution is $dx=0.09$ heliocentric degrees, {\it i.e.} $dx=1.093$ Mm (the HMI resolution is degraded, but this does not introduce significant aliasing in the power spectra), and the temporal cadence is $dt=45$ s.
A typical tracked cube has the following format: $384 \times 384 \times 7681$, where the first two dimensions are the spatial coordinates, and the last one is time.
We select $62.5$ h from each cube (most cubes have a few bad frames to be rejected).
We set  $R_\mathrm{min}=25.15$ Mm, for all the sunspots to be contained inside the inner disk (except sunspot NOAA 11263, the largest studied which extends beyond $R_\mathrm{min}$ and is strongly asymmetrical). The outer radius of the annulus is $R_\mathrm{max}=209$ Mm. There is a limit to the maximum azimuthal order $m$ at which $A_m$ and $B_m$ can be calculated: The inner-circle circumference $2 \pi R_\mathrm{min}= 158$ Mm corresponds to $\approx 144$ sampling points. Therefore, $m$ (the number of nodes along the circle) should not exceed $144/2=72$. Here we somewhat arbitrarily limit $m$ to $-46 \le m \le 46$, because there is not much oscillation signal at higher $m$.
The resolution in wavenumber ($k$) is $dk=2\pi/(R_\mathrm{max}-R_\mathrm{min})=0.0342$ Mm$^{-1}$, {\it i.e.} the resolution in angular degree $\ell$ is equal to $d\ell=dk \times R_{\odot}=23.8$ (where $R_{\odot}=696$ Mm is the solar radius).
The resolution in temporal frequency ($\nu$) is $d\nu=0.00444$ mHz.
Figure \ref{firstfig} shows the geometry used for the Hankel-Fourier analysis with respect to sunspot NOAA 11289.
Table \ref{table.sunspots} lists all of the 15 sunspots studied.

\begin{table}[t]
\caption{List of sunspots studied.}\label{table.sunspots}
\begin{tabular}{ccc}
\hline
NOAA number & Start time (TAI) & Stop time (TAI)\\[3pt]
\hline
11092 & 2010.07.31\_12:00:00 & 2010.08.04\_12:00:00 \\
11243 & 2011.07.01\_20:00:00 & 2011.07.05\_20:00:00 \\
11263 & 2011.08.01\_12:00:00 & 2011.08.05\_12:00:00 \\
11289 & 2011.09.10\_00:00:00 & 2011.09.14\_00:00:00 \\
11306 & 2011.09.29\_00:00:00 & 2011.10.03\_00:00:00 \\
11314 & 2011.10.13\_12:00:00 & 2011.10.17\_12:00:00 \\
11317 & 2011.10.14\_00:00:00 & 2011.10.18\_00:00:00 \\
11352 & 2011.11.19\_00:00:00 & 2011.11.23\_00:00:00 \\
11384 & 2011.12.24\_00:00:00 & 2011.12.28\_00:00:00 \\
11388 & 2011.12.31\_00:00:00 & 2012.01.04\_00:00:00 \\
11408 & 2012.01.25\_00:00:00 & 2012.01.29\_00:00:00 \\
11410 & 2012.01.30\_00:00:00 & 2012.02.03\_00:00:00 \\
11419 & 2012.02.16\_12:00:00 & 2012.02.20\_12:00:00 \\
11420 & 2012.02.16\_17:00:00 & 2012.02.20\_17:00:00 \\
11423 & 2012.02.27\_12:00:00 & 2012.03.02\_12:00:00 \\
\hline
\end{tabular}
\end{table}

\begin{figure}
\centering
\includegraphics[width=\textwidth]{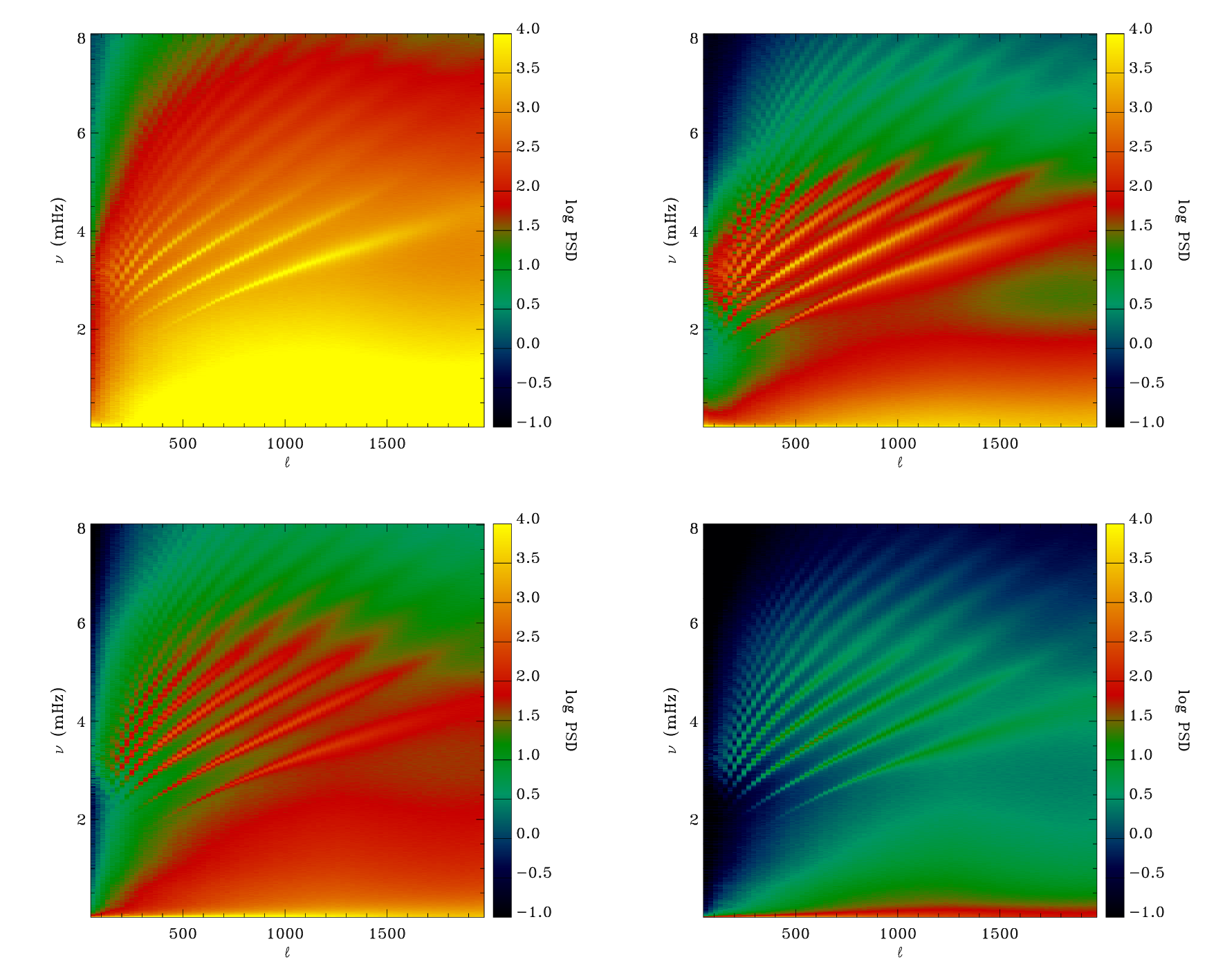}
\caption{Maps of the power spectral density as a function of angular degree $\ell$ and temporal frequency $\nu$, averaged over four quiet-Sun regions in the HMI continuum intensity (upper left panel), HMI Dopplergram (upper right panel), AIA 1700 \AA\ intensity (lower left panel), and AIA 1600 \AA\ intensity (lower right panel). The color scale is truncated.}
\label{firstfigb}       
\end{figure}

Figure \ref{firstfigb} shows the PSDs (averaged over four quiet-Sun regions) for ingoing plus outgoing waves, as a function of $\ell$ and $\nu$ and for the HMI continuum, Doppler velocity, AIA 1700 \AA, and AIA 1600 \AA\ intensity data. The power spectra are clearly different. In particular, the high power in the granulation domain ($\nu \le 1.8$ mHz) on the HMI-continuum data is conspicuous, but it has already significantly dropped at the Doppler-velocity level.

\section{Results}

Four quiet-Sun regions are studied besides 15 sunspots. The quiet Sun was initially envisioned merely as a control of the Hankel-Fourier transform code, but it turned out to present some new and interesting features on its own, and will be discussed after the sunspot results.

\subsection{Outgoing-Wave Power-Reduction and Enhancement in Sunspots}

\begin{figure}
\centering
\includegraphics[width=\textwidth]{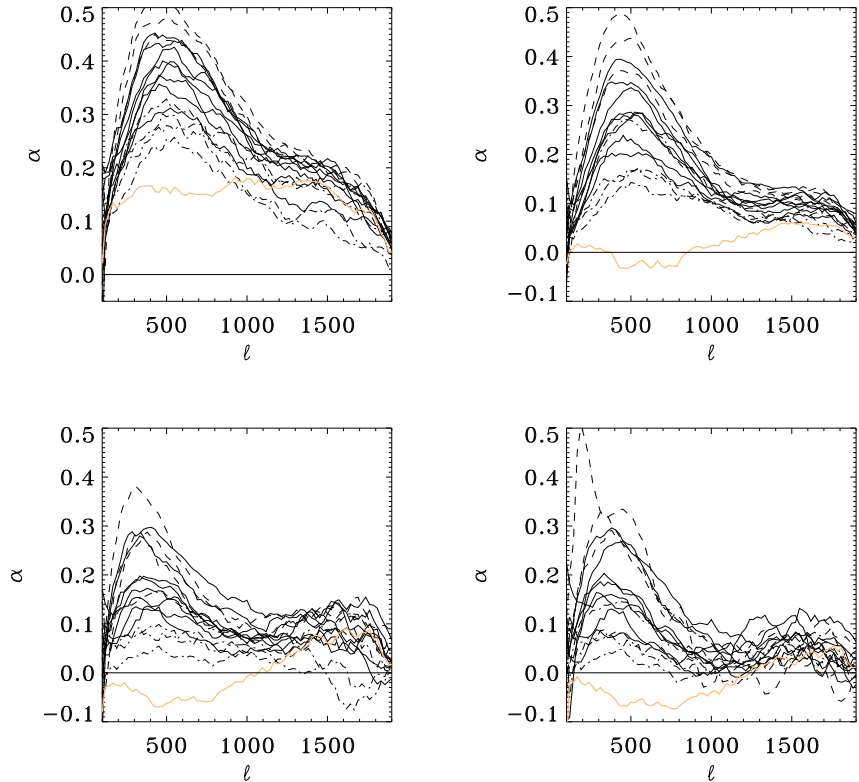}
\caption{Power-reduction coefficient $\alpha(k)$ as a function of angular degree $\ell$ for 15 sunspots in the HMI continuum intensity (upper left panel), HMI Dopplergram (upper right panel), AIA 1700 \AA\ intensity (lower left panel), and AIA 1600 \AA\ intensity (lower right panel). The orange curves are from a quiet-Sun region, the dashed lines are for sunspots with a LOS magnetic flux inside the disk of radius $R_\mathrm{min}$ larger than $8.4 \times 10^{21}$ Mx, the dash-dotted lines are for a flux lower than $2.1 \times 10^{21}$ Mx, and the solid black lines are for fluxes in-between. Background power is removed and an $f$ and $p$-mode mask is applied. The curves are smoothed.}
\label{secfig}       
\end{figure}

\begin{figure}
\centering
\includegraphics[width=\textwidth]{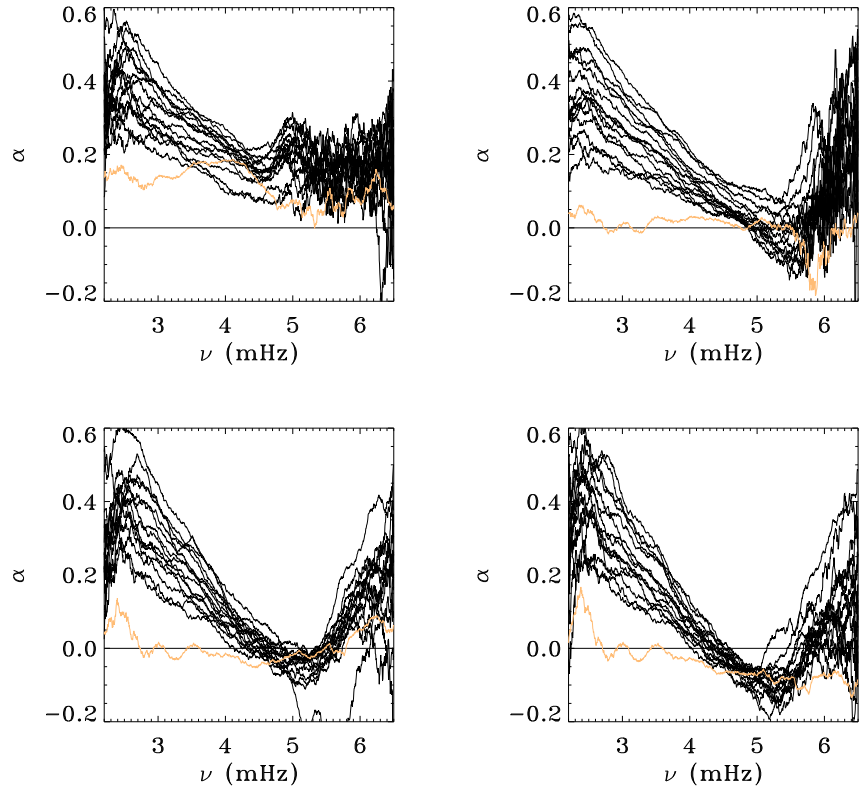}
\caption{Power-reduction coefficient $\alpha(\nu)$ as a function of temporal frequency $\nu$ for 15 sunspots in the HMI continuum intensity (upper left panel), HMI Dopplergram (upper right panel), AIA 1700 \AA\ intensity (lower left panel), and AIA 1600 \AA\ intensity (lower right panel). Background power is removed and an $f$ and $p$-mode mask is applied. The curves are smoothed. The orange curves are for a quiet-Sun region.}
\label{thirdfig}       
\end{figure}

To facilitate the comparison with past articles on power reduction in sunspots, we first calculate $\alpha(k)$ (Figure \ref{secfig}) and $\alpha(\omega)$ (Figure \ref{thirdfig}). A background power is fitted to the $m$-averaged ingoing-plus-outgoing spectra, and removed. An $f$ and $p$-mode mask selects the $f$ and $p_1$ to $p_9$ ridges above 2 mHz for the computation of $\alpha(k)$ and $\alpha(\omega)$. 

The different panels of Figure \ref{secfig} look similar to the upper panel of Figure 10 of Bogdan {\it et al.} (1993). Overall, sunspots do reduce the power of outgoing waves ($\alpha(k) > 0$). There is an increase in $\alpha(k)$ with $\ell$, for $\ell \le 500$, and then a decrease for larger $\ell$. The peak power reduction occurs at the HMI continuum level and reaches $0.52$. This is comparable to numbers quoted by Braun, Duvall, and LaBonte (1988) and larger than those of Bogdan {\it et al.} (1993). Any difference may be due to the sunspots themselves, to the way the background power is subtracted, to the $f$ and $p$-mode mask applied, and to the higher $m$ reached ({\it e.g.}, Braun, Duvall, and LaBonte (1988) stopped at $m=5$, Bogdan {\it et al.} (1993) stopped at $m=15$, and Braun (1995) stopped at $m=20$).

 The panels of Figure \ref{secfig}, characterizing four different heights in the photosphere, display different $\alpha(k)$ values, implying a dependence on height. The angular degree $\ell$ of peak power reduction slightly decreases with height from $\ell \approx 500$ at the continuum level to $\ell \approx 400$ higher in the photosphere (and even to lower $\ell$ for NOAA 11263 in AIA 1600 \AA\ data). 

On Figure \ref{thirdfig}, $\alpha(\omega)$ peaks below $\nu=3$ mHz, and decreases toward higher frequencies. The steepness of this decrease with $\nu$, increases with atmospheric height for roughly $\nu < 4.5$ mHz. Moreover, a power reduction at the continuum level in the frequency range $\nu \approx 4.5-5.5$ mHz turns into a power enhancement ($\alpha(\omega) < 0$) in the upper photosphere. The existence of a negative $\alpha(\omega)$ in this frequency range and for sunspot data has not been widely reported in previous publications. Braun (1995) mentioned a drop close to zero in power absorption near the photospheric acoustic cut-off frequency ($\nu \approx 5.3$ mHz), and an increase at higher frequency, but no enhancement (although Figure 8 of Braun (1995) shows some amount of ``emission'' in the ranges $\ell=329-390$ and $\ell=411-473$, with only two sunspots analyzed). To the best of our knowledge, only Chen {\it et al.} (1996) explicitly mentioned a power enhancement, using K-line data formed in the upper photosphere/lower chromosphere from the {\it Taiwan Oscillation Network}. With their annulus closest to the sunspot and similar to the one ($R_\mathrm{min}$,$R_\mathrm{max}$) used here, this power enhancement appears even at frequencies lower than in the present study, at least for one of their sunspots. 

The height dependence of $\alpha(\omega)$ for waves with $\nu \le 5.3$ mHz is difficult to explain if we assume that solar oscillations below the acoustic cut-off frequency (in the quiet Sun) are purely evanescent. However, propagation at frequencies below this cut-off was observed in sunspot penumbrae (McIntosh and Jefferies, 2006; Rajaguru {\it et al.}, 2007). Even in quiet Sun, Canfield and Musman (1973) early on discussed about low-frequency propagating waves. Roberts (1983) described how dissipation in the atmosphere lowers the cut-off frequency. Because radiative losses are important at the height of the HMI-continuum formation, there might not be a clear division between propagating and evanescent oscillations. More recently, De Pontieu, Erdelyi, and James (2004) investigated how inclined fields allow $p$-mode leakage, and Khomenko {\it et al.} (2008) showed how the 5-min oscillations can propagate through the photosphere to the chromosphere in small vertical magnetic-flux tubes. Stodilka (2010) showed that granulation lowers the cut-off frequency in the solar photosphere, allowing for upward leakage of oscillation power.
Finally, Howe {\it et al.} (2012) studied the phase of acoustic waves in and around active regions, using HMI and AIA observables. A phase difference (implying propagation) in regions surrounding these sunspots is visible between HMI Dopplergram and AIA 1600 \AA\ intensity. All of these results, and the present paper, favor the idea that solar oscillations below the acoustic cut-off frequency are not purely evanescent. 
 
\begin{figure}
\centering
\includegraphics[width=\textwidth]{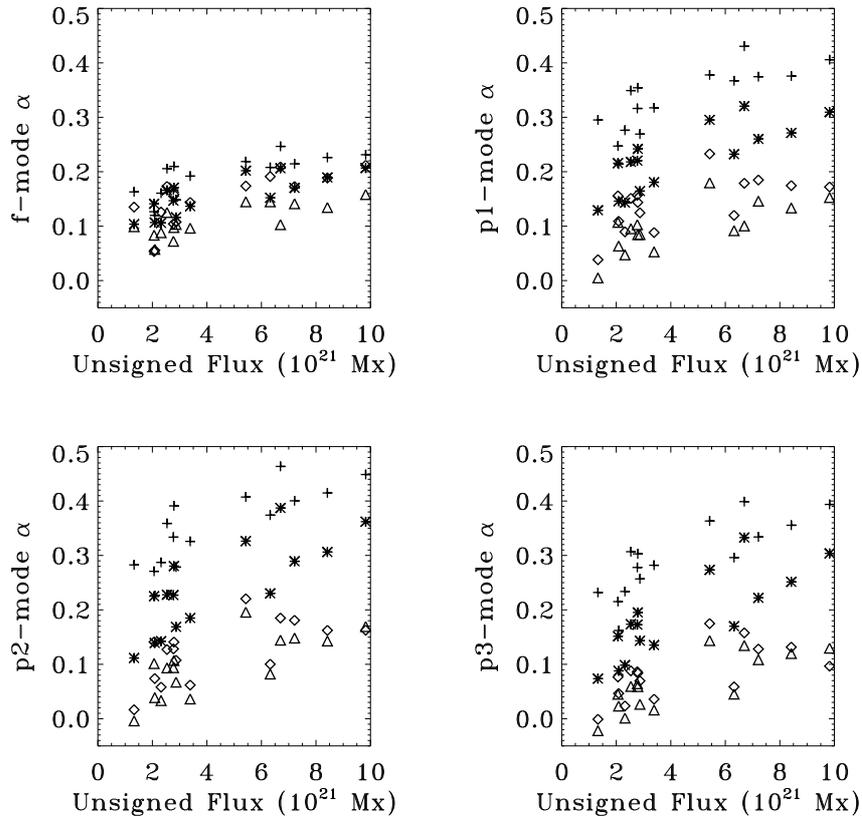}
\caption{Power-reduction coefficient $\alpha$ for the sunspots in the HMI continuum intensity (crosses), HMI Dopplergram (stars), AIA 1700 \AA\ intensity (diamonds), and AIA 1600 \AA\ intensity (triangles). Upper left panel is for the $f$ modes, upper right panel is for the $p_1$-modes, lower left panel is for the $p_2$-modes, and lower right panel is for the $p_3$-modes. Background power is subtracted.}
\label{fifthfigb}       
\end{figure}

\begin{figure}
\centering
\includegraphics[width=\textwidth]{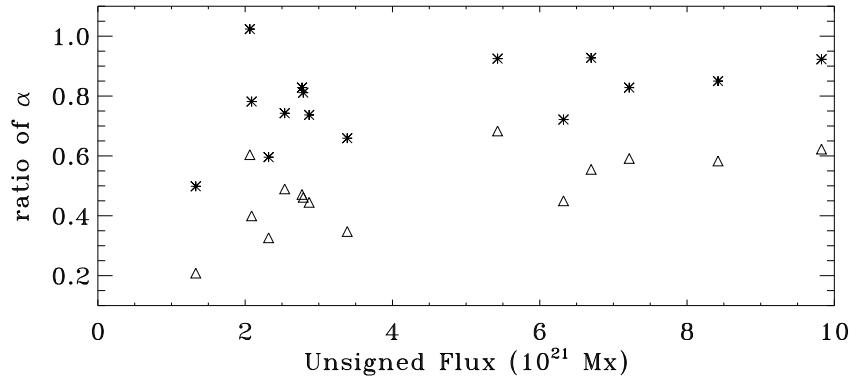}
\caption{Ratio of power reduction ($\alpha$) in HMI Dopplergram to power reduction in continuum intensity (stars), and ratio of power reduction in AIA 1600 \AA\ intensity to power reduction in continuum intensity (triangle), as a function of the unsigned LOS magnetic flux of sunspots. Background power is subtracted and an $f$ and $p$-mode mask is applied. Only the $f$ and $p$-modes in the frequency range $\nu < 4.5$ mHz are selected.}
\label{fifthfigbbb}       
\end{figure}

Figure \ref{fifthfigb} shows $\alpha$ as a function of the unsigned LOS magnetic flux inside the disk of radius $R_\mathrm{min}$, and for separate radial orders ($n$) of the oscillation modes (background power is removed). $\alpha$ is maximum at continuum level, and decreases with height: perhaps an indication that the mechanism responsible for power reduction operates mainly near, or below, continuum level. Maximum reduction occurs for $p_1$ and $p_2$-modes, reduction is weaker for $p_3$-modes, and minimum reduction occurs for $f$-modes. There is a dependence of $\alpha$ on magnetic flux: Larger fluxes produce greater reduction. This dependence seems stronger for $p$-modes than for $f$-modes. Saturation might be reached for large fluxes, but additional sunspots are needed to increase the sample size and access stronger fields. On Figure \ref{fifthfigbbb}, the ratios of $\alpha$ in HMI Dopplergram to $\alpha$ in HMI continuum, and of $\alpha$ in AIA 1600 \AA\ to $\alpha$ in HMI continuum, both appear to increase with magnetic flux (although there is significant scatter in the data points). Thus, the region of outgoing-wave power reduction could extend higher in the photosphere as the flux increases.
The main contributor to this power reduction is considered to be mode conversion ({\it e.g.} Spruit and Bogdan, 1992; Cally and Bogdan, 1993; Crouch {\it et al.}, 2005; Khomenko and Collados, 2006), where $p$-modes suffer absorption by partial conversion to MAG and Alfv\'en waves. The slow MAG modes travel upward toward the chromosphere, along magnetic-field lines, while fast MAG modes travel upwards but are progressively refracted and eventually reflected, probably in the lower chromosphere. Mode conversion occurs primarily in the equipartition layer, where Alfv\'en speed equals sound speed.

\begin{figure}
\centering
\includegraphics[width=\textwidth]{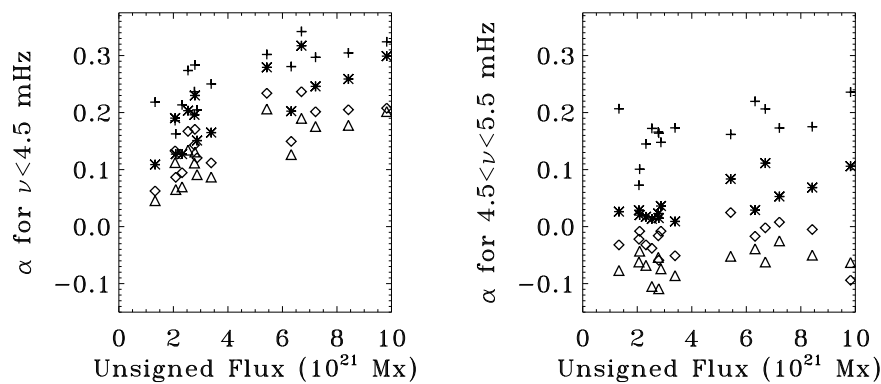}
\caption{Power-reduction coefficient ($\alpha$) for the sunspots in the HMI continuum intensity (crosses), HMI Dopplergram (stars), AIA 1700 \AA\ intensity (diamonds), and AIA 1600 \AA\ intensity (triangles). Left panel is for the $f$ and $p$-modes in the frequency range $\nu < 4.5$ mHz, right panel is for the $f$ and $p$-modes in the frequency range $4.5<\nu<5.5$ mHz. Background power is subtracted and an $f$ and $p$-mode mask is applied.}
\label{fifthfigbb}       
\end{figure}

Because some oscillation modes have their power enhanced by sunspots (Figure \ref{thirdfig}), the frequency ranges $\nu<4.5$ mHz and $4.5<\nu<5.5$ mHz are separated in Figure \ref{fifthfigbb}. This ensures that power enhancement for modes in the range $\nu \approx 4.5-5.5$ mHz does not impact power-reduction estimates at lower frequencies. The power enhancement is mainly present in AIA data (triangles and diamonds), {\it i.e.} the upper photosphere and lower chromosphere (although some sunspots also show a weaker enhancement in Doppler data), and seems to only have a weak dependence on the LOS magnetic flux.

\begin{figure}
\centering
\includegraphics[width=\textwidth]{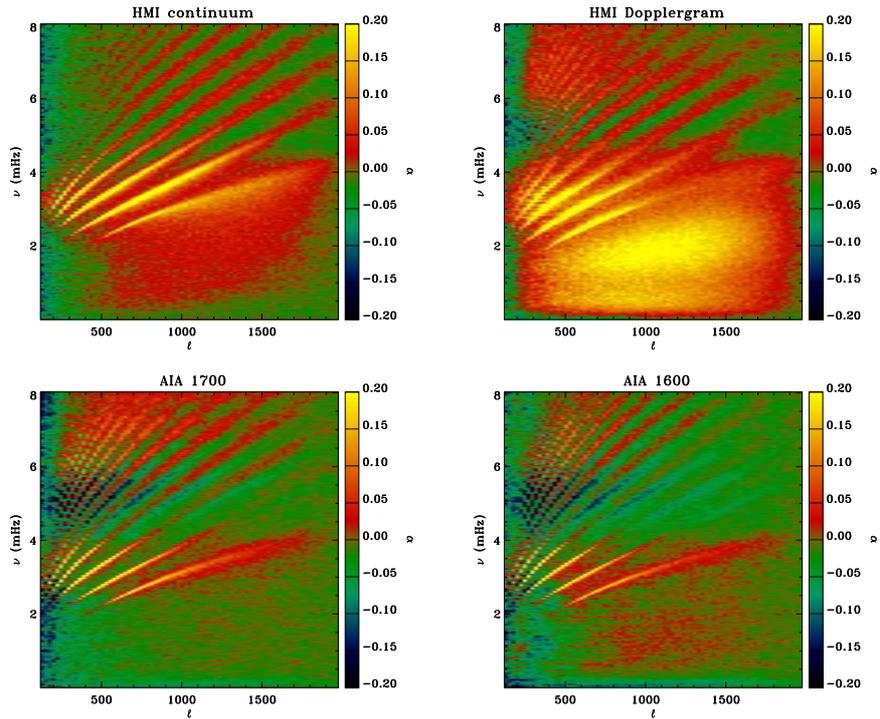}
\caption{Maps of power-reduction coefficient $\alpha(k,\omega)$ as a function of angular degree $\ell$ and temporal frequency $\nu$ averaged over 15 sunspots in the HMI continuum intensity (upper left panel), HMI Dopplergram (upper right panel), AIA 1700 \AA\ intensity (lower left panel), and AIA 1600 \AA\ intensity (lower right panel). The color scale is truncated to $-0.2$ to $+0.2$. The maps are smoothed to reduce the noise level.}
\label{fourthfig}       
\end{figure}

Due to the long duration of the tracked cubes, the high resolution and excellent quality of HMI and AIA data, it is possible to present the oscillation-power reduction/enhancement in sunspots in a more informative fashion by drawing 2D maps of $\alpha(k,\omega)$, as was done on Figure 5 of Bogdan {\it et al.} (1993) with lower resolution and a higher noise level.
To increase the S/N ratio, the maps of $\alpha(k,\omega)$ for the 15 sunspots are averaged all together (despite each sunspot having a different size, shape, and magnetic flux). The result is shown in Figure \ref{fourthfig}. No background subtraction is performed. Therefore, $\alpha(k,\omega)$ in the oscillation-mode ridges is underestimated.
These maps present some striking features. First, the power enhancement mostly seen with AIA data on Figure \ref{thirdfig} and in the range $\nu \approx 4.5-5.5$ mHz is conspicuous. It is probably a manifestation of acoustic halos (Brown {\it et al.}, 1992; Braun {\it et al.}, 1992). The term acoustic glory (Braun and Lindsey, 1999; Donea, Lindsey, and Braun, 2000) is also used when referring to these extended halos where acoustic power surrounding a sunspot or a plage is enhanced. 

\begin{figure}
\centering
\includegraphics[width=\textwidth]{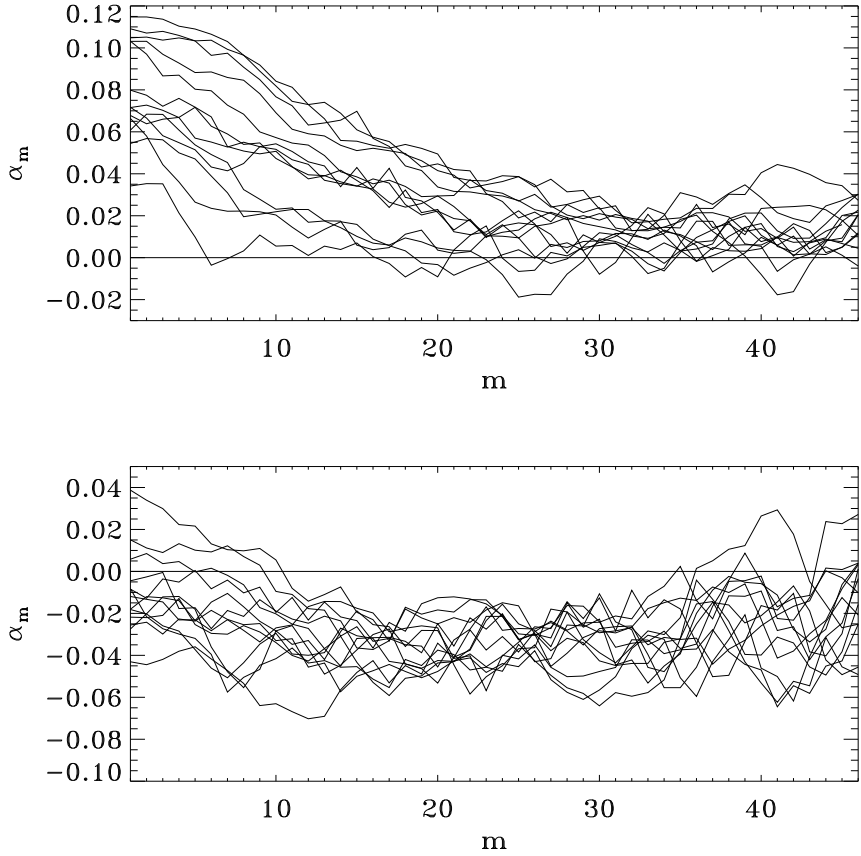}
\caption{Power-reduction coefficient $\alpha_m$ as a function of $m$ for the sunspots in the AIA 1600 \AA\ channel. Upper panel shows the power reduction integrated over all $k$ and $\nu$ (after applying the $f$ and $p$-mode mask). Lower panel shows the power reduction in the frequency range $\nu = 4.5-5.5$ mHz (after applying the $f$ and $p$-mode mask). The curves are smoothed to reduce the noise level. No background power is removed.}
\label{fifthfig}       
\end{figure}

To try and determine where this enhancement occurs, Figure \ref{fifthfig} shows the individual power-reduction coefficients ($\alpha_m$) obtained for the AIA 1600 \AA\ data as a function of $m$, for $f$ and $p$-mode ridges at all frequencies in the upper panel, and for $p$-mode ridges in the frequency range $\nu=4.5-5.5$ mHz in the lower panel (with mask applied to select $n \le 9$, but no background removed).
Plotting $\alpha_m$ as a function of azimuthal order $m$ provides information about the spatial extent of the power-reduction/enhancement region. Indeed, as Braun, Duvall, and LaBonte (1988) point out, the ``impact parameter'' of a mode increases with $m$: The energy density of an incident plane wave falls off exponentially for a radial distance $r<m/k$. In other words, the higher $m$ is, the less sensitive a mode is to the central part (small $r$) and the more sensitive it becomes to the peripheral part (large $r$) of the region under consideration.
Therefore, in the upper panel of Figure \ref{fifthfig}, a drop in $\alpha_m$ with increasing $m$ is consistent with the power-reduction region being contained inside the inner disk of radius $R_\mathrm{min}$. $\alpha_m$ levels off at roughly $m \ge 30$ but does not reach zero for some sunspots, perhaps due to the presence of plages or other magnetized regions outside these sunspots: The region of power reduction seems to extend beyond the sunspots. Indeed, most sunspots are surrounded by plages visible in AIA data (for example, see Figure \ref{firstfig}), and plages are also present further inside the annulus where the Hankel-Fourier transform is performed (contaminating our results).
In the lower panel of Figure \ref{fifthfig} ($\nu=4.5-5.5$ mHz) the power enhancement occurs mostly at $m \ge 10$. At lower $m$, the enhancement is less significant or even turns into a power reduction for some sunspots. This may imply that this enhancement occurs not toward the center, but rather toward the edges of --- or even outside --- sunspots. This supports the idea that it is a manifestation of acoustic glories.
Moreover, this enhancement is not present at the continuum level (upper left panel of Figure \ref{fourthfig}), in agreement with the observation that there is no acoustic glories at this level ({\it e.g.} Hindman and Brown, 1998; Jain and Haber, 2002; Howe {\it et al.}, 2012). Acoustic glories are explained by Khomenko and Collados (2009) as resulting from fast MAG waves being progressively refracted as they propagate upward in the sunspot. Because of this refraction, the waves escape the active region.
Another possibility is that halos result from scattering of waves by the magnetic field of sunspots ({\it e.g.} Hanasoge, 2009).

\begin{figure}
\centering
\includegraphics[width=\textwidth]{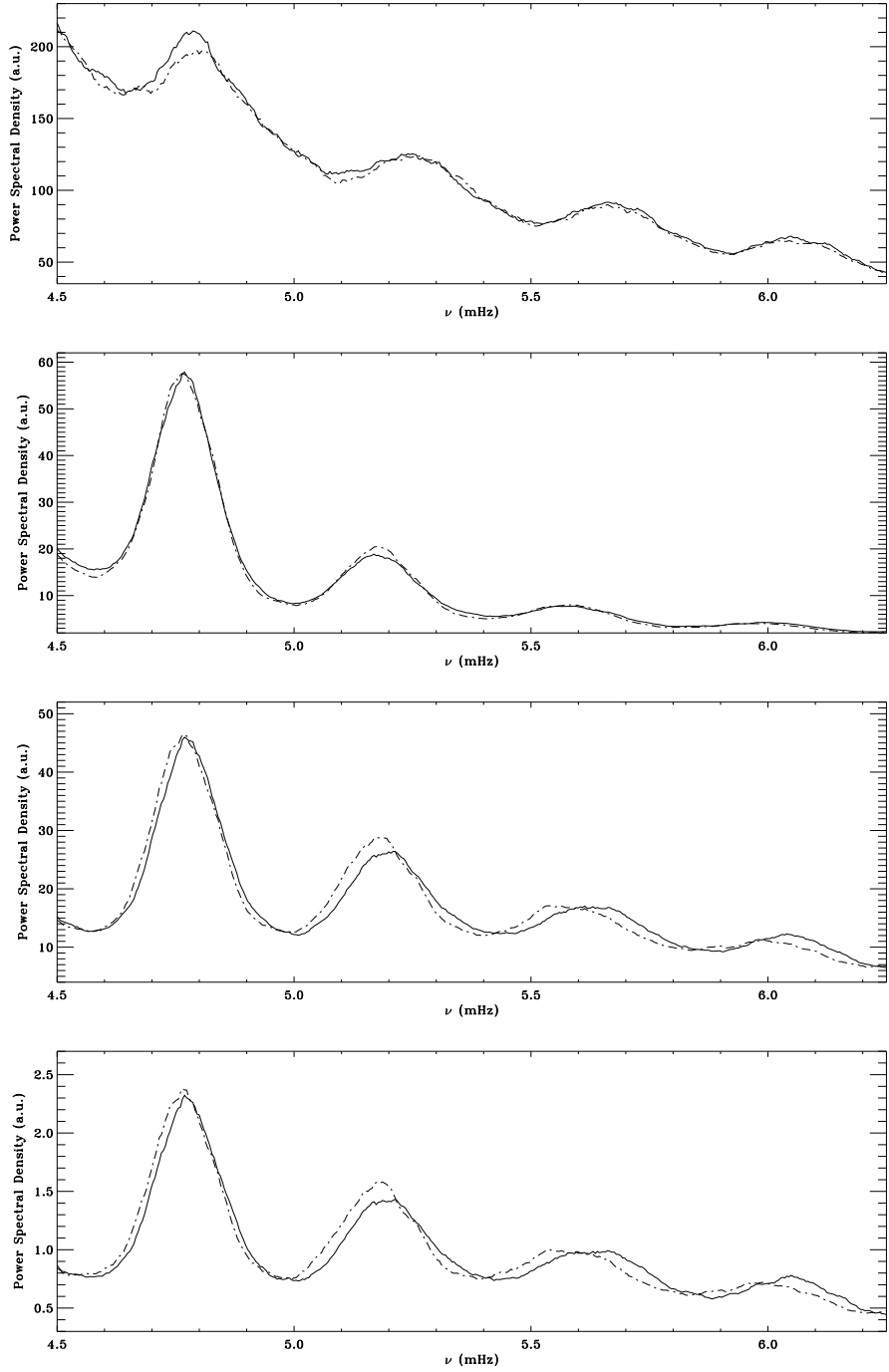}
\caption{Power spectral density of the ingoing (solid lines) and outgoing (dash-dotted lines) waves as a function of the temporal frequency $\nu$, averaged for the 15 sunspots, for continuum intensity data (upper panel), Dopplergrams (second row from top), AIA 1700\AA\ (third row from top), and AIA 1600 \AA\ (lower panel). These power spectra are cuts at $\ell=357$. The curves are smoothed.} 
\label{fifthfige}       
\end{figure}

\begin{figure}
\centering
\includegraphics[width=\textwidth]{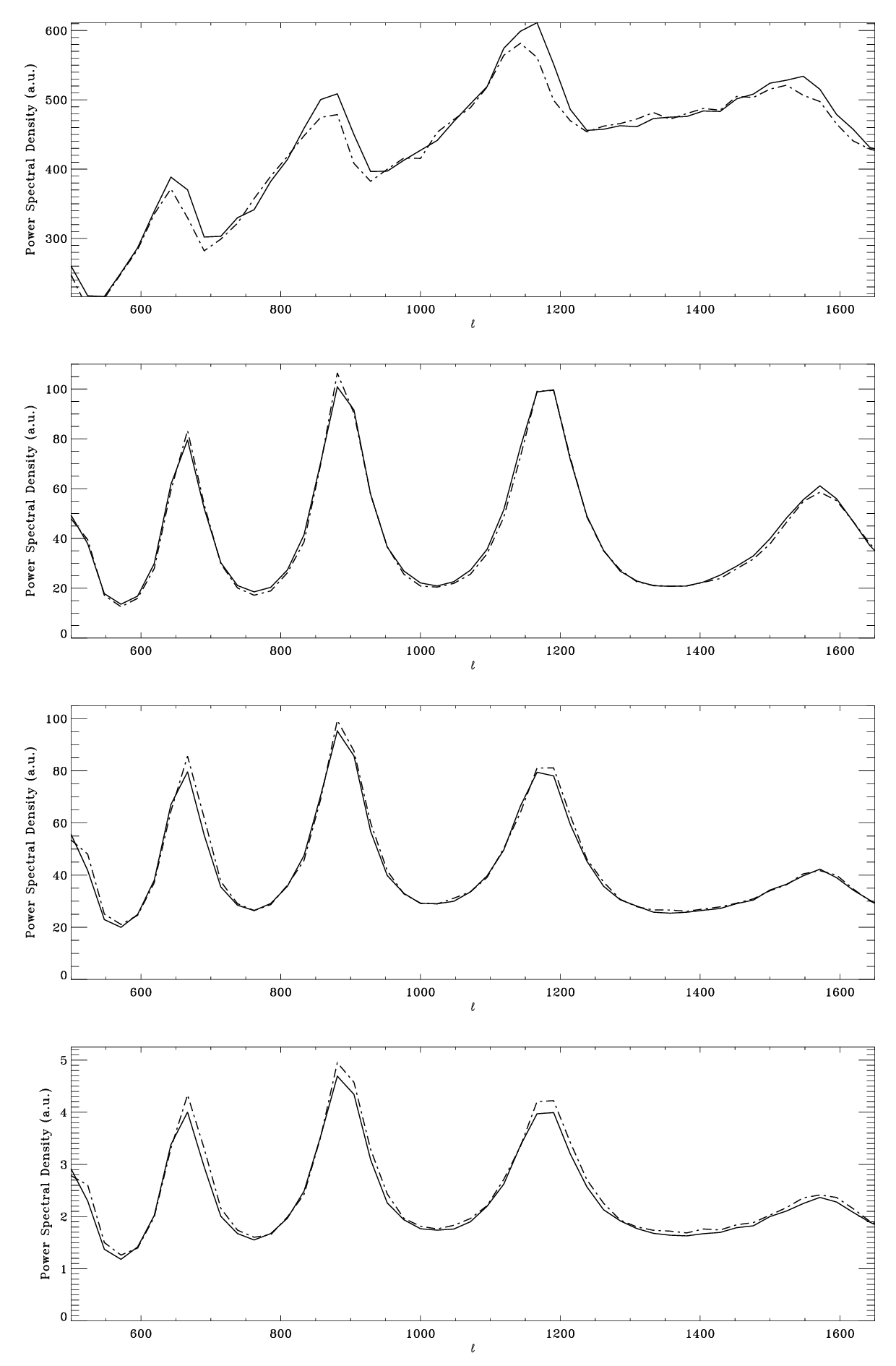}
\caption{Power spectral density of the ingoing (solid lines) and outgoing (dash-dotted lines) waves as a function of the angular degree $\ell$, averaged for the 15 sunspots, for continuum intensity data (upper panel), Dopplergrams (second row from top), AIA 1700\AA\ (third row from top), and AIA 1600 \AA\ (lower panel). These power spectra are cuts at $\nu=5$ mHz. The curves are smoothed.} 
\label{fifthfigf}       
\end{figure}


Figures \ref{fifthfige} and \ref{fifthfigf} show cuts in the average PSDs of ingoing and outgoing waves (average over the 15 spectra from sunspot data), and as a function of $\nu$ (Figure \ref{fifthfige}) and $\ell$ (Figure \ref{fifthfigf}).
In the lower panel of Figure \ref{fifthfige} (cut at $\ell=357$ for the AIA 1600 \AA\ data), the frequencies ($\nu$) of outgoing $p$-modes at $\nu \ge 4.5$ mHz ($p_6$ to $p_9$-modes) are reduced compared to ingoing modes. This frequency shift $d\nu$ increases with $\nu$. A simple Gaussian fit of individual mode peaks returns a frequency difference (outgoing minus ingoing) of $d\nu=-0.0098$ mHz for $p_6$, $d\nu=-0.0209$ mHz for $p_7$, and $d\nu=-0.054$ mHz for $p_8$.

Such an effect is absent (or less significant) at lower frequencies, and instead we obtain $d\nu=+0.0095$ mHz for $p_1$-modes at $\ell=357$. At $\ell \approx 1400$, $d\nu$ of $p_1$-modes is close to zero. Finally, the $f$-modes seem to always show a small frequency increase ($d\nu = 0.0104$ mHz at $\ell = 357$ and $d\nu = 0.007$ mHz at $\ell = 1400$) for outgoing waves. These results contradict Braun, Duvall, and LaBonte (1988) who observed no difference (within error bars) in the ridge positions of the ingoing and outgoing waves. The decrease in frequency of outgoing waves at higher $\nu$ is absent at continuum or Doppler velocity level, and is visible only on AIA data. On the other hand, there could be a small positive $d\nu$ at all frequencies in the HMI continuum (for $p_1$ at $\ell=357$, $d\nu=0.003$ mHz, {\it i.e.} below the resolution in $\nu$).
 Frequency increase in AIA data for low-$\nu$ modes is compatible with the results of Sun {\it et al.} (1997), who did observe such an increase on K-line images. They interpreted it as the effect of outgoing flows in and around sunspots. However, their Figure 1 seems to show a positive $d\nu$ for $\nu > 4.5$ mHz, where we observe a negative one. The origin of the frequency decrease of higher-$\nu$ modes in AIA data is more difficult to ascertain.

This negative $d\nu$ on AIA data partly contributes to the existence of a negative $\alpha(\omega)$ in the range $\nu \approx 4.5-5.5$ mHz, but the lower panel of Figure \ref{fifthfige} and the Gaussian fits of power peaks confirm that the power enhancement is genuine: The PSD is indeed larger for outgoing waves. Incidentally, the peak at $\nu \approx 5.2$ mHz highlights that a power enhancement of outgoing waves is present even in Doppler data. Therefore, this enhancement develops rather low in the photosphere and becomes stronger with height.

Power reduction/enhancement could also partly be due to wave scattering by inhomogeneities in sunspots (Braun, Duvall, and LaBonte, 1987), and not only to actual power absorption/emission: Ingoing waves may be scattered into higher wavenumbers. Braun, Duvall, and LaBonte (1987) and Hanasoge (2009) consider such a change in wavelength at fixed frequency. Scattering of incoming waves by sunspots also affects their phases, {\it e.g.} Braun (1995). Figure \ref{fifthfigf} shows cuts in the power spectra at $\nu=5$ mHz and as a function of $\ell$. In AIA data, a Gaussian fit of individual mode peaks finds a very small angular-degree shift $d\ell$ of outgoing waves toward higher $\ell$, but the resolution in $\ell$ is not fine enough to confirm its existence ($d\ell=1.934$ at $\ell=890$ and $\nu=5$ mHz, and $d\ell=2.23$ at $\ell=1179$, well below the resolution in $\ell$).
In any case, a small shift in wavelength and/or frequency of outgoing waves is not the primary driver of the power enhancement observed in the upper photosphere in the $\nu \approx 4.5-5.5$ mHz range, and of the power reduction observed at all heights at lower frequencies. 

Noticeably, power-reduction coefficients calculated from Figure \ref{fifthfigf} are higher (by $\approx 0.01$ at continuum level) than those shown on Figure \ref{fourthfig}. This difference arises because the average of the ratios outgoing PSD to ingoing PSD is not equal to the ratio of the average of outgoing PSDs to  the average of ingoing PSDs.

\begin{figure}
\centering
\includegraphics[width=\textwidth]{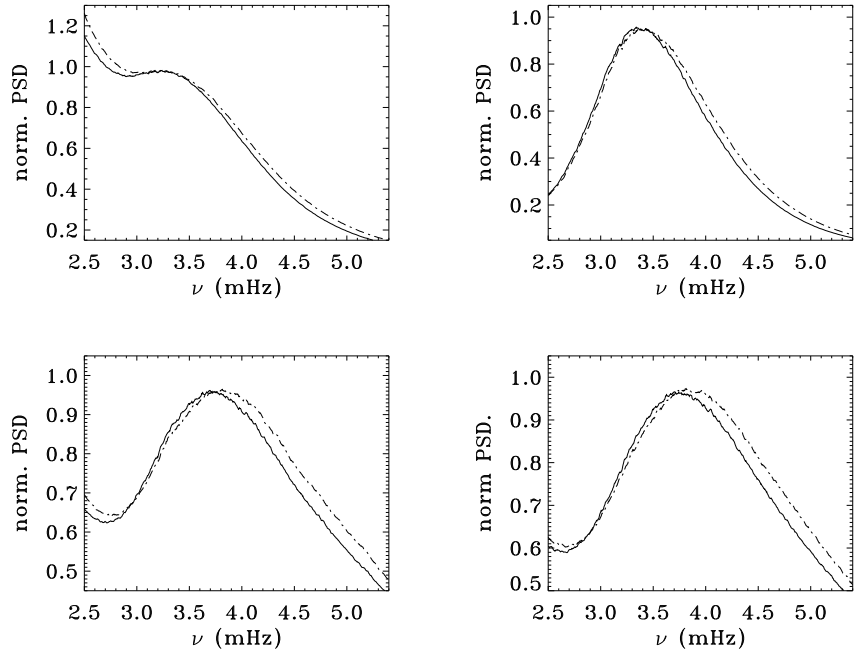}
\caption{Power spectral density of the ingoing (solid lines) and outgoing (dash-dotted lines) waves normalized by their maximum values, as a function of the temporal frequency $\nu$, averaged for the 15 sunspots, for continuum intensity data (upper left panel), Dopplergrams (upper right panel), AIA 1700 \AA\ (lower left panel), and AIA 1600 \AA\ (lower right panel). }
\label{fifthfigg}       
\end{figure}

On Figure \ref{fifthfigg}, the 15 PSDs from sunspot data are averaged all together over $k$ and normalized by their respective maximum values, to compare the frequency at which maximum power occurs. With AIA 1600 \AA\ the peak power of outgoing waves is shifted by $d\nu = 0.087$ mHz compared to ingoing waves (a Gaussian fit is used). At AIA 1700 \AA\ level, the shift is slightly smaller ($d\nu = 0.082$ mHz), and at Doppler-velocity level it is only $d\nu = 0.040$ mHz. There is no significant shift at the continuum level. Therefore, $d\nu$ is altitude dependent and results from the sunspots, since a similar plot with quiet-Sun data shows $d\nu \approx 0$ at all heights. Again, Sun {\it et al.} (1997) interpreted this overall frequency increase in the upper photosphere as resulting from outflows in sunspots.
It is also noteworthy that while the oscillation power peaks at $\nu \approx 3.4$ mHz with HMI continuum and velocity data, it peaks at $\nu \approx 3.8$ mHz with AIA data: probably reflecting the sensitivity to chromospheric signal where the preferred oscillation period is 3 min instead of 5 min at the photospheric level.

\begin{figure}
\centering
\includegraphics[width=\textwidth]{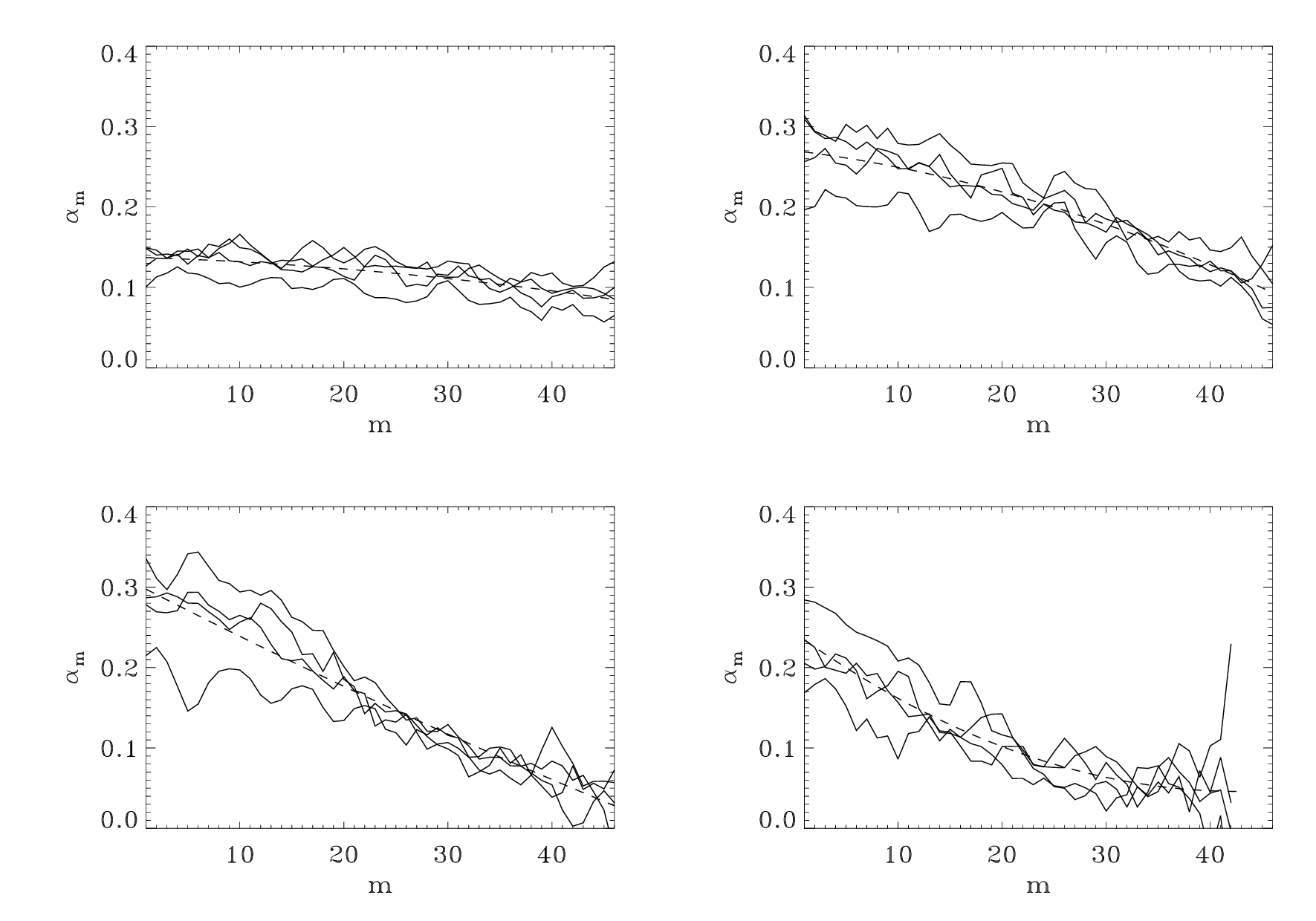}
\caption{Power-reduction coefficient $\alpha_m$ for four sunspots in the HMI continuum intensity as a function of the azimuthal order $m$. Background power is not subtracted. The upper left panel is for the $f$-modes only, the upper right panel is for the $p_1$-modes, the lower left panel is for the $p_2$-modes, and the lower right panel is for the $p_3$-modes. The dashed lines are a second-order polynomial fit to the data.}
\label{fifthfigc}       
\end{figure}

Figure \ref{fifthfigc} shows $\alpha_m$ as a function of $m$, for the $f$, $p_1$, $p_2$, and $p_3$-modes separately, and at the HMI continuum level (only for the first four sunspots of Table \ref{table.sunspots} to make the figure more easily readable).
$\alpha_m$ decreases in all cases with $m$, but this decrease is steeper for $p$-modes than $f$-modes. Under the assumption that this dependence on $m$ mainly reflects a difference in spatial location where power reduction occurs (but could also partly be due to, {\it e.g.}, a lack of symmetry of the absorption region about the polar axis), it appears that at the continuum level the outgoing-wave $p$-mode power is reduced mainly toward the sunspot center while the outgoing-wave $f$-mode power is reduced more evenly across the entire inner disk (and maybe even outside this disk). Indeed, $f$-modes show a smaller drop in $\alpha_m$ as $m$ increases. For the $p_3$-modes, $\alpha_m$ levels off at $m \approx 30$.

\begin{figure}
\centering
\includegraphics[width=\textwidth]{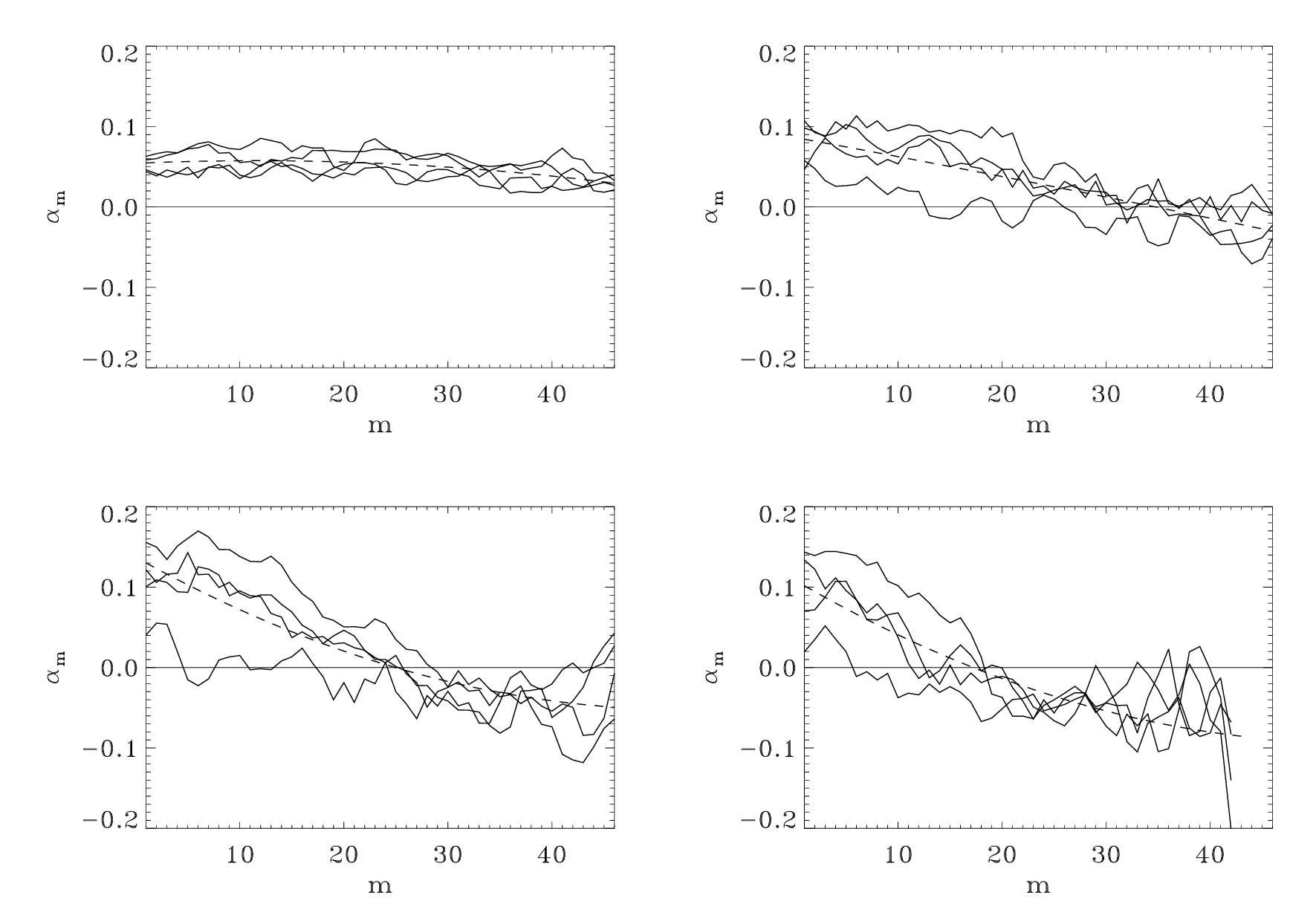}
\caption{Power-reduction coefficient $\alpha_m$ for four sunspots in the AIA 1600 \AA\ intensity as a function of the azimuthal order $m$. Background power is not subtracted. The upper left panel is for the $f$-modes only, the upper right panel is for the $p_1$-modes, the lower left panel is for the $p_2$-modes, and the lower right panel is for the $p_3$-modes. The dashed lines are a second-order polynomial fit to the data.}
\label{fifthfigcc}       
\end{figure}

Figure \ref{fifthfigcc} is the same as Figure \ref{fifthfigc}, but at AIA 1600 \AA\ level. Here too, the decrease in $\alpha_m$ is steeper for $p$-modes than $f$-modes, and this power reduction turns into power enhancement for $p$-modes as $m$ increases (only higher $\ell$ are accessible when $m$ increases because the impact parameter of the waves increases: The power spectra become restricted to high-$\ell$ high-$\nu$ modes).

\begin{figure}
\centering
\includegraphics[width=\textwidth]{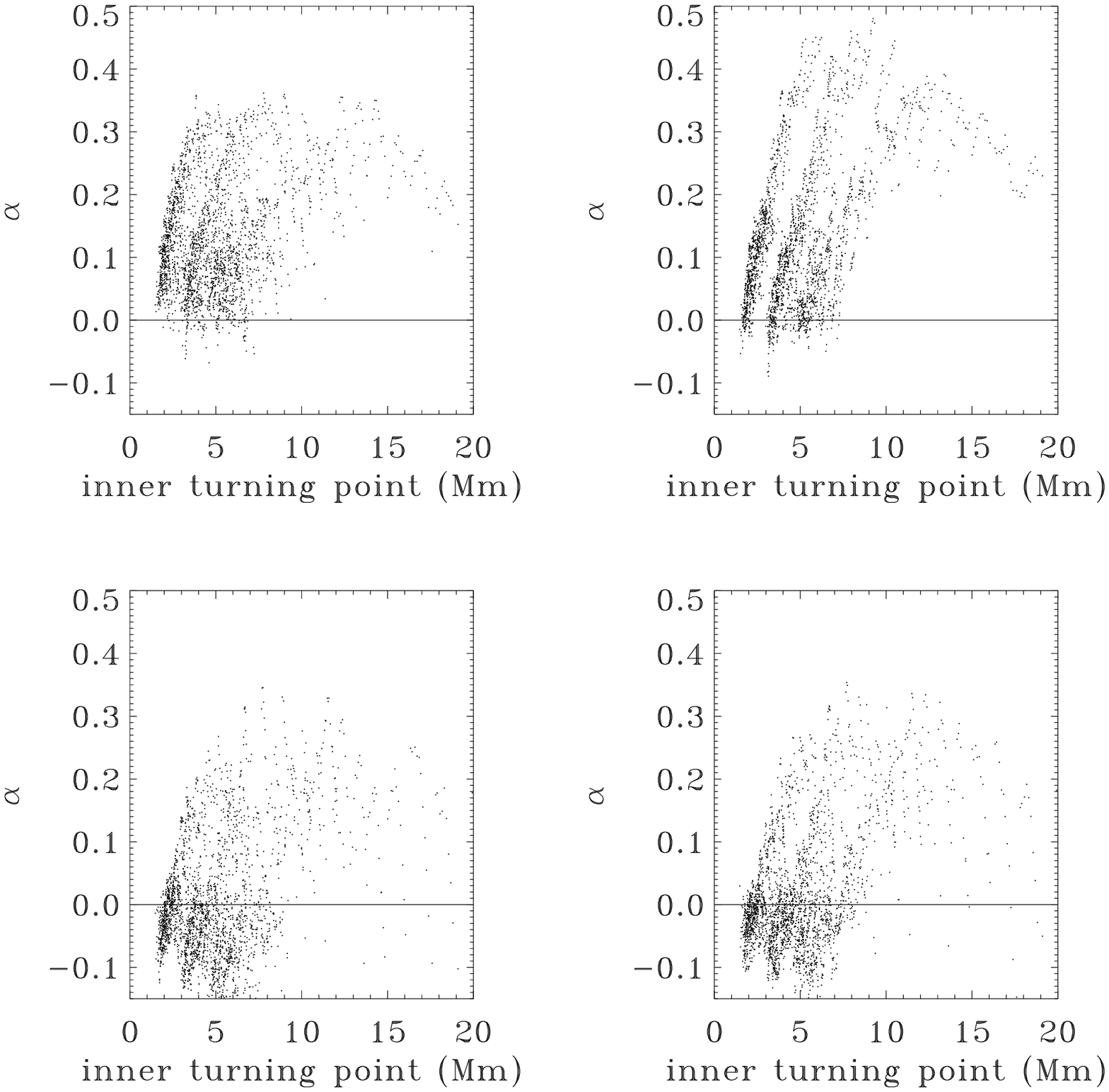}
\caption{Power-reduction coefficient $\alpha$ for sunspot NOAA 11289 as a function of inner turning point of the $p_1$ (leftmost curves), $p_2$, and $p_3$ (rightmost curves) modes. Background power is not subtracted. Shown are the HMI continuum intensity (upper left panel), HMI Dopplergram (upper right panel), AIA 1700 \AA\ (lower left panel), and AIA 1600 \AA\ (lower right panel). The curves are smoothed.} 
\label{fifthfigd}       
\end{figure}

Finally, Figure \ref{fifthfigd} shows $\alpha$ as a function of the inner turning points $R_0$ of modes, calculated in the ray-path approximation, for sunspot NOAA 11289. Only the first three radial orders $n$ are plotted. $R_0$ is defined as the depth at which $c/R_0=\omega / \ell$, where $c=c(r)$ is the sound speed. Solar model S of \mbox{J.} Christensen-Dalsgaard provided $c(r)$.
There is a dependence of $\alpha$ on $R_0$ that is similar for the three radial orders studied.
$\alpha$ first increases with $R_0$, then it reaches a plateau, and it eventually decreases. This behaviour could be explained if the power reduction occurs in a layer extending underneath the photosphere. However, for $n=3$ and at the continuum level, $\alpha$ increases until $R_0 \approx 14$ Mm. At such a depth, the magnetic pressure is so low compared to gas pressure that it is unlikely that the field of sunspots has any impact. Therefore, the dependence of $\alpha$ on $R_0$ may actually be a dependence on another parameter correlated with $R_0$: {\it e.g.}, the angle of incidence of waves reaching their upper turning point varies with $R_0$, and the inclination of $p$-modes with respect to magnetic-field lines impacts mode conversion. For AIA data, the power enhancement in each ridge occurs for the modes with the shallowest lower turning point, {\it i.e.} the modes closest to the surface.

\subsection{Outgoing-Wave Power-Reduction and Enhancement in Quiet Sun}

On Figures \ref{secfig} and \ref{thirdfig}, $\alpha(k)$ and $\alpha(\omega)$ are plotted for a quiet-Sun region (in orange). HMI Dopplergrams and AIA data show a weak power enhancement peaking around $\ell \approx 600$, as observed first by Bogdan {\it et al.} (1993) and later confirmed by Braun (1995) and Chen {\it et al.} (1996), and a weak power reduction at higher $\ell$.
Bogdan {\it et al.} (1993) explained this enhancement by assuming that there is a statistical equilibrium at the solar surface where power absorption and emission are balanced. Modes that are truly global in nature have their power reduced when crossing a sunspot or a pore somewhere on the solar surface, and this will create a power emission at some other location in the quiet Sun where the Hankel-Fourier transform is performed. Chen {\it et al.} (1996) also mentioned the existence of a center-to-limb effect, an excess of power for waves propagating from Sun center toward the limb, contributing to a negative $\alpha$.

However, HMI continuum data on Figure \ref{secfig} show a different picture. There is a nearly constant and positive $\alpha(k)$ in the range $\ell \approx 200-1500$, with an amplitude of $\alpha(k) \approx 15$\% (when background power is subtracted). This somewhat puzzling result, opposite to what happens higher in the photosphere, was confirmed by running the Hankel-Fourier code on three other quiet-Sun regions observed at different dates and latitudes. We also ran a different code provided by Douglas \mbox{C.} Braun (2012, private communication), and performed the Hankel-Fourier analysis on a tracked cube of full-disk MDI continuum-intensity data to confirm that HMI and MDI continuums behave consistently.
Finally, as a test, we processed HMI intensity images prior to the Hankel transform analysis: The limb darkening was (mostly) removed and a running difference of the images was taken (to get rid of low-frequency granulation signal). This pre-processing had only a minor impact on $\alpha$.
 
\begin{figure}
\centering
\includegraphics[width=\textwidth]{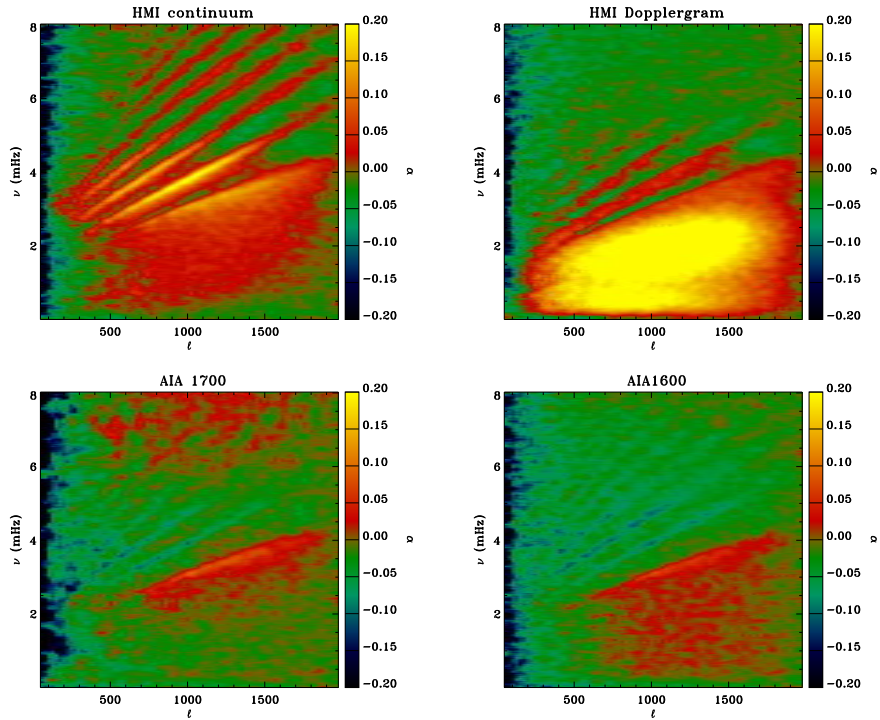}
\caption{Maps of power-reduction coefficient $\alpha(k,\omega)$ as a function of angular degree $\ell$ and temporal frequency $\nu$, averaged over four quiet-Sun regions in the HMI continuum intensity (upper left panel), HMI Dopplergram (upper right panel), AIA 1700 \AA\ (lower left panel), and AIA 1600 \AA\ (lower right panel). The color scale is truncated to $-0.2$ to $+0.2$.}
\label{sixthfig}       
\end{figure}

Figure \ref{sixthfig} shows the outgoing-wave power reduction $\alpha(k,\omega)$ maps averaged over four quiet-Sun regions. The continuum-intensity map stands out as being the only one showing a systematic power reduction in the $f$ and $p$-mode ridges. Other maps show a residual reduction in the power of $f$-modes, but no reduction and even a weak power enhancement in some $p$-mode ridges.
The source of strong power reduction at the continuum level is unknown, but it is suggested by Thomas \mbox{L.} Duvall (2012, private communication) that granulation plays a role. Indeed, Hankel-Fourier analysis performed on HMI line-core intensities of the four quiet-Sun regions shows no significant power reduction in the $p$-mode range. The main difference between continuum and line-core intensities is a $\approx 200$ km shift in the height of formation, resulting in significantly less granulation signal at the line-core level.
The existence of this ubiquitous power reduction at the continuum level seems to invalidate the assumption in Bogdan {\it et al.} (1993) regarding a statistical balance between emission and absorption at the solar surface, when height is not taken into account.

Kostyk and Shchukina (1999), Khomenko {\it et al.} (2001), and Kostyk, Shchukina, and Khomenko (2006) showed that the amplitude of the 5-min oscillations is much larger over intergranules than over granules and that oscillations are suppressed over granules. Roth {\it et al.} (2010) confirmed that waves excited in the intergranular lanes lose energy while probing a granule. Kostyk, Shchukina, and Khomenko (2006) found that even though there are waves propagating upwards over both granules and supergranules, the waves over granules are closer to standing waves than those over intergranular lanes.
These results might partly explain the positive $\alpha(k,\omega)$ obtained with the Hankel-Fourier analysis. Indeed, numerous granules are present inside the inner disk studied. With a typical diameter of 1 Mm, $\approx 630$ granules are expected at any time in the disk of radius $R_\mathrm{min}$. Acoustic waves excited outside this disk and traveling toward its center might have their power reduced when crossing the central part of those granules. However, this does not explain why this power reduction is asymmetrical between ingoing and outgoing waves.

A plot of $\alpha$ as a function of mode lower turning point ($R_0$) shows no dependence (or only a weak one) on $R_0$ for Doppler velocity and AIA data, as expected.
However, at the continuum level, $\alpha$ increases with $R_0$ up to $R_0 \approx 6.5$ Mm for $n=3$ (the data points are very scattered). This might indicate that the mechanism responsible for power reduction at the continuum level operates in a shallow layer underneath the solar surface, maybe related to the typical height of a granule.

 Another conspicuous feature of Figure \ref{sixthfig} is the strong power reduction below the $f$-mode ridge on HMI Dopplergrams, absent, or weak, on intensity maps. This power reduction reaches maximum below $\nu=2$ mHz and peaks at the angular degree $\ell \approx 950$ (wavelength of $\approx 4.5$ Mm) corresponding to granular flows ({\it e.g.} Hathaway {\it et al.}, 2000). Therefore, in Doppler signal and unlike in intensity, the quiet Sun reduces the power of outgoing granular flows. The reason is currently unknown, but is probably somehow related to wave properties specific to Doppler measurements, \mbox{vs.} intensity measurements. This power reduction is stronger in quiet Sun than in sunspots (see Figure \ref{fourthfig}), where granulation is inhibited by strong magnetic fields.


Finally, it was suggested that we run the Hankel-Fourier analysis on numerical simulations of the upper convective zone (\mbox{T.L.} Duvall, Jr., 2012, private communication) to try and understand the origin of the power reduction in $f$ and $p$-modes in the continuum data, and in the granular-flow domain in Doppler data. To this end, the 96 Mm-wide MHD simulations of Robert Stein were used ({\it e.g.}, Stein and Nordlund, 2000; Benson, Stein, and Nordlund, 2006): Data cubes of intensity at the solar surface can be downloaded from http://steinr.pa.msu.edu/$\sim$bob/data.html, with a cadence of $dt = 60$ s and a spatial resolution of $dx \approx 0.095$ Mm. About 8 h of data are available. Partly because of the lower duration of the dataset, and mostly because of its relatively small size ($96 \times 96$ Mm$^2$), the Hankel-Fourier coefficients are noisy. There does not seem to be any power reduction for outgoing oscillation modes present in the simulation, at least not above noise level (rms variation for $\alpha(k,\omega)$ is $\approx 0.18$), despite the conspicuous presence of granulation signal. Similarly, the Hankel-Fourier analysis run on simulated Doppler-velocity data finds no outgoing-wave power reduction in the granulation domain (or it is below noise level: The rms variation for $\alpha(k,\omega)$ in Doppler data is also $\approx 0.18$).

\section{Conclusion}

High temporal-cadence and high-resolution data taken simultaneously at different wavelengths by the HMI and AIA instruments onboard the SDO satellite make it possible to seismically probe the lower layers of the solar atmosphere and to study wave propagation and properties from the continuum level to the lower chromosphere. A Hankel-Fourier analysis was applied to 15 sunspots and four quiet-Sun regions observed by SDO. As was already known, sunspots reduce the power of oscillation modes crossing them at all heights in the atmosphere for frequencies lower than $\nu \approx 4.5$ mHz. This is likely a result of mode conversion by magnetic fields, reduced emissivity in the sunspots, mode scattering, etc. This power reduction increases with the LOS magnetic flux. The outgoing-wave power reduction seems to be stronger at the continuum level than higher in the atmosphere (when background power is subtracted prior to computing $\alpha$), perhaps indicating that the source of reduction is located low in --- or even below --- the photosphere. The height dependence of the outgoing-wave power-reduction coefficient ($\alpha$) also favors the idea that part of the solar oscillations below the acoustic cut-off frequency (in the quiet Sun) do propagate upward.
On sunspot data, we observe a power enhancement higher in the photosphere for waves in the frequency range $\nu \approx 4.5-5.5$ mHz. This enhancement is barely visible in Doppler-velocity data, but is strong at the AIA 1700 and 1600 \AA\ level. Such a negative $\alpha$ was reported by Chen {\it et al.} (1996).
This is likely the signature of acoustic glories observed around sunspots. A weaker power enhancement is also present in quiet-Sun data of the upper photosphere, as was reported by Bogdan {\it et al.} (1993) and Chen {\it et al.} (1996). Sunspots also shift the frequencies of outgoing waves, depending on the altitude and the angular degrees and frequencies of the modes. The frequency increase observed for some modes in AIA data is similar to the results of Sun {\it et al.} (1997), interpreted as evidence of outflows in sunspots.

 In the quiet Sun, there is a striking dependence of the power-reduction coefficient on height. At the continuum level, the quiet Sun behaves like a power reducer for $p$ and $f$-oscillation modes. This puzzling behaviour, never reported before as far as we know, is suggested by \mbox{Thomas L. Duvall, Jr.} (2012, private communication), to be somehow related to the presence of granulation. Indeed, granulation power is prominent in HMI continuum data but quickly vanishes higher in the atmosphere. Closer to the chromosphere, the quiet Sun becomes a power enhancer in most $p$-mode bands, while continuing to absorb $f$-modes. In Doppler-velocity signal, there is a strong outgoing-wave power reduction in the domain of granular flows. A weaker reduction is also present in sunspot data.

Hankel-Fourier analysis performed on Doppler velocity and continuum data from one of \mbox{R.} Stein's upper-convective-zone MHD simulation shows no power reduction (above noise level) at the continuum level in the $f$ and $p$-mode bands, and no power reduction in Doppler data at the surface in the granular-flow domain. These simulations do have a clear granulation signal though, proving that the relations between granulation and the observed power reduction in the HMI continuum, and granulation and the power reduction in granular flows at Doppler-velocity level, are not straightforward. 

This mostly observational study leaves many questions unanswered. Future work will focus on the origin of the power reduction of $f$ and $p$-modes in the quiet Sun at the continuum level (and the power enhancement higher), and on the reduction in granulation power in Doppler data. We also plan to compute the phase differences between ingoing and outgoing waves, to gather information about their running or standing nature, and about scattering phenomena. More sunspots need to be analyzed, to access larger magnetic fluxes. Hankel-Fourier analysis should also be performed on line-core intensity data in a systematic way, to access yet another height in the photosphere, and to emphasize the difference between intensity and Doppler-velocity signals (as line-core and Doppler signals are formed at similar altitudes). Several seismic studies have recently been performed combining HMI and AIA data, and a comparison of their results is warranted: this could help interpret some of the results obtained here.

\begin{acks}
This work was supported by NASA Grant NAS5-02139 (HMI). The data used here are courtesy of NASA/SDO and the HMI and AIA science teams. We thank Douglas \mbox{C.} Braun for providing his Hankel-Fourier analysis codes, and Thomas \mbox{L.} Duvall, Jr., for his help, numerous advice, and suggestions regarding the role of granulation in our results. We also thank Robert \mbox{F.} Stein for providing his 96 Mm-wide numerical simulation results. Finally, we are grateful to the anonymous referee whose suggestions improved this paper.
\end{acks}

\end{article}
\end{document}